\documentclass{webofc}
\usepackage[varg]{txfonts}   
\def \arcsec{\hbox{$^{\prime\prime}$}\,}
\def \arcmin{\hbox{$^{\prime}$}\,}
\def \mic{$\mu$m}
\def \micb{$\mu$m }
\def \deg{$^\circ$}
\begin{document}
\title{Dust evolution in pre-stellar cores}
\subtitle{To appear in the proceedings of the international conference entitled mm Universe @ NIKA2, Grenoble (France), June 2019, EPJ Web of conferences }
%

\author{\firstname{Charlène} \lastname{Lefèvre}\inst{1,2}\fnsep\thanks{\email{lefevre@iram.fr}} \and 
	\firstname{Laurent} \lastname{Pagani}\inst{2} 
	\and
	\firstname{Bilal} \lastname{Ladjelate}\inst{3} 
	\and
	\firstname{Michiel} \lastname{Min}\inst{4}
	\and
	\firstname{Hiroyuki} \lastname{Hirashita}\inst{5}
	\and
	\firstname{Robert} \lastname{Zylka}\inst{1}
}

	\institute{ Institut de Radioastronomie Millim\'etrique
	(IRAM), 300 rue de la Piscine, 38400 Saint-Martin
	d'H\`eres, France
	\and
	{LERMA \& UMR8112 du CNRS, Observatoire de Paris, PSL Research University, CNRS, Sorbonne Universités, UPMC Univ. Paris 06,  F- 75014 Paris, France}
	\and
	Instituto de Radioastronomía Milimétrica, IRAM, Avenida Divina Pastora 7, Local 20, E-18012, Granada, Spain
	\and
	Sterrenkundig Instituut Anton Pannekoek, University of Amsterdam, Science Park 904, 1098 XH Amsterdam, The Netherlands
	\and
	Institute of Astronomy and Astrophysics, {Academia Sinica, Astronomy--Mathematics Building, AS/NTU, No.1, Sec. 4, Roosevelt Road,} Taipei 10617, Taiwan
}

\abstract{%
  Dust grains are the building {blocks} of future planets. They evolve in size, shape and composition during the life cycle of the interstellar medium. We seek to understand the process which leads from diffuse medium grains to dust grains in the vicinity of protostars inside disks. As a first step, we propose to characterize the dust evolution inside pre-stellar cores thanks to multi-wavelength observations. We will present how NIKA2 maps are crucial to better constrain dust properties and {we will} introduce SIGMA: a new flexible dust model in open access.}
\maketitle
\section{Introduction}
\label{intro}

Dust evolution starts at the earliest stage of star formation, during the formation of cores that slowly contract to form pre-stellar cores (PSC), to the collapse of these PSCs into protostars and protoplanetary disks (PPD). {The contraction time of PSCs is still a debated question and ranges from less than \,1\,My \cite{2013Pagani,2017Kortgen} to 10\,My lifetime \cite{2006Mouschovias}.} {PSCs are dense ($>$10$^5$\,H\,.\,cm$^{-3}$), compact (<\,10$^4$\,AU) and cold (5--12 K) objects, making their} study difficult. {They are very opaque and} most gaseous species are depleted onto grains preventing the study of the inner parts. Dust is the only tracer that is present from cloud edge to the densest part, allowing to characterize cloud density structure. Nevertheless, dust itself is a poor tracer in visible and near-infrared (NIR) since its absorption is too high to detect the reddening of the stars above A$\mathrm{_V}$\,$\sim$\,50\,mag. Recent observations of PSCs with Spitzer \cite{2010Pagani,2014Lefevre,2016Lefevre} and PPDs with SPHERE \cite{2017SPHERE} in scattered light {has brought} hope to put more constraints on dust properties. Indeed, all parameters involved in the dust emission (dust temperature, density, grain size, emissivity and spectral index) are varying across the cloud leading to degenerate solutions when dust emission is used alone \cite{2015Pagani}. Studies combining dust scattering and emission reproduced successfully PPD observations, allowing to constrain the geometry and density structure in PPDs \cite[e.g.][]{2019Villenave}. On the other hand, a consistent multi-wavelength modeling of PSCs remains an unachieved goal. Grain emissivity in the far--infrared (FIR) has been linked consistently to its absorption efficiency in the near--infrared (NIR) at short wavelengths only \cite{2015aJuvela} or at low resolution with 5\arcmin Planck data \cite{2015bJuvela}, and thus never reproducing observations of the densest part. With our on-going NIKA2 open time Program, we aim at building a consistent multi-wavelength picture of two neighbor molecular clouds, L183 and L134 {(116$\pm$6 pc [C. Zucker priv. comm.] and 107$\pm$5 pc \cite[]{2019Zucker} respectively)} hosting 4 PSCs.

\section{Method}
\label{intro}

\subsection{Constraints obtained from other wavelengths}

 The first step of our approach is to build a 3D density model of the cloud independently from scattered light or degenerated emission process. Our density model relies on dust extinction as measured by reddening in the NIR, and the PSC densities are deduced from molecular observations of N$_2$H$^+$ and N$_2$D$^+$ \cite{2007Pagani,2016Lefevre}, assuming thermalized gas and dust and rotational symmetry\footnote{Though simplistic hypothesis, some limits on the clumpiness of the PSC can also be obtained from self-absorbed scattering, as observed for L183.} (see Fig.~1).

 \begin{figure}[h]
 	\begin{center}
 		\begin{minipage}[b]{0.7\linewidth}
 			\includegraphics[scale=0.4]{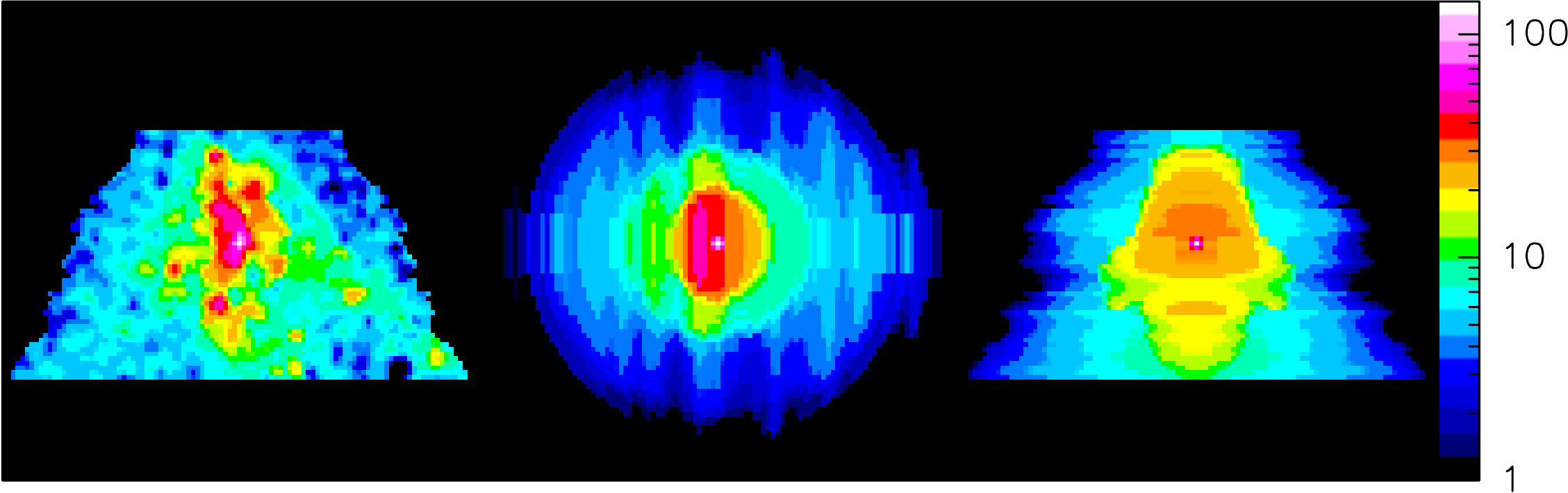}
 		\end{minipage}
 		\begin{minipage}[b]{0.2\linewidth}
 			\includegraphics[scale=0.35]{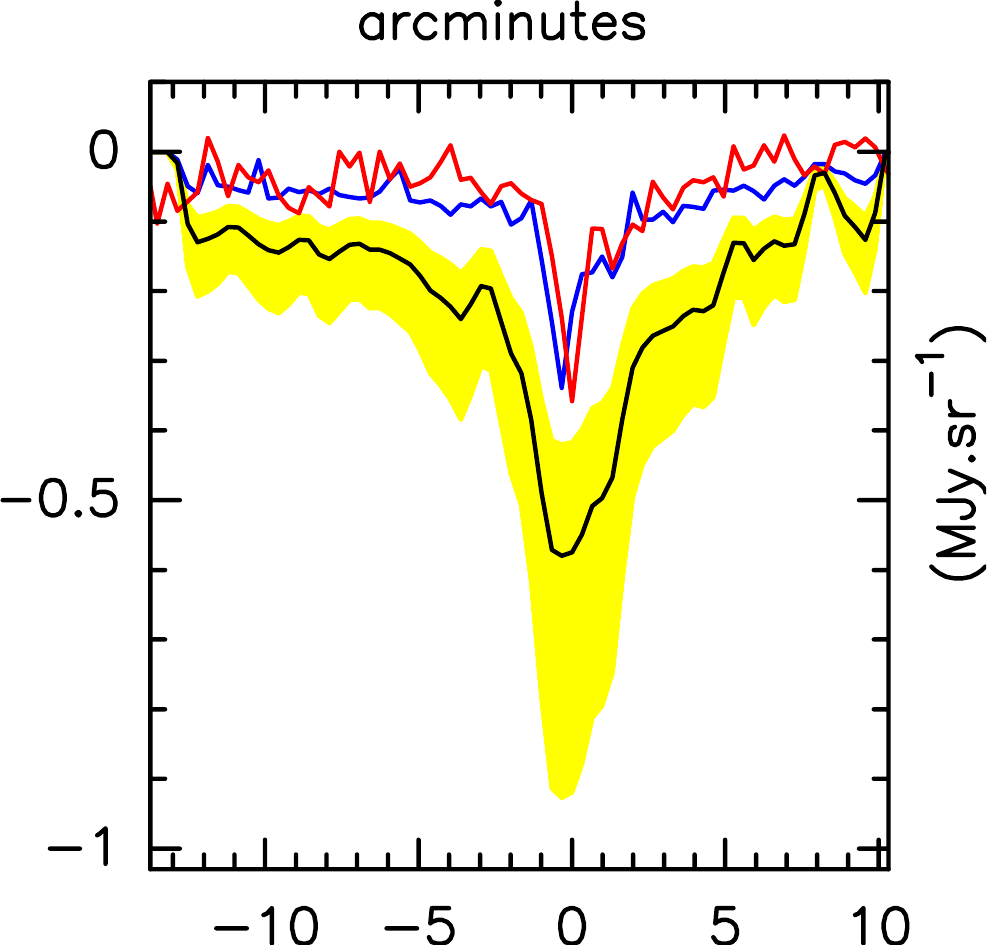}
 		\end{minipage}
 		\label{fig-cloud_model}       
 		\caption{A$_\mathrm{V}$ maps of L183 with color scale in magnitudes. The 3D density model is based on NIR extinction of background stars and N$_2$H$^+$ and N$_2$D$^+$ observations for the PSC \cite{2016Lefevre}. From left to right: facing the cloud, from above, from the side. \textbf{Right panel:} 8\,\micb cut profile through the PSC with observations in red, modeling in blue, and modeling without the scattering contribution in black with uncertainties from background estimation in yellow.}
 	\end{center}
 \end{figure}
 
  Regarding dust properties, we rely on mid-infrared (MIR) scattering to constrain dust geometries (shape and sizes).  In particular, to reproduce MIR observations of L183  we demonstrate that an important fraction of large aggregates with an equivalent size of 4\,\micb \cite{2016Min} is requested, in particular to explain the scattering contribution remaining at 8\,\micb \cite[][and Fig.~1 right]{2016Lefevre}. Though the grain model we proposed has the suitable geometry to explain MIR observations, its emissivity is too high in the FIR/mm (see Fig.~2). While the dust size and porosity are well constrained by MIR scattering (called coreshine), the dust emissivity strongly varies depending on their composition. Recent laboratory measurements show the impact of the iron fraction inside silicates on dust emissivity \cite{2017aDemyk,2017bDemyk}. Here, the silicates included in the dust mixture are pure enstatite (MgSiO3, 75\%) mixed with iron sulfides (FeS, 10\%) and carbonaceous (15\%). We developed a code to test the impact of dust composition, in particular silicates and ice mantles, on dust emissivities. SIGMA is an open access code\footnote{\url{https://github.com/charlenelefevre/SIGMA}} to compute the dust properties of icy aggregates from NIR to mm wavelengths based on laboratory measurements (see Sect.~\ref{sec-SIGMA}). While we have already found a reasonable solution to model L183 in the NIR and MIR \cite{2016Lefevre}, the NIKA2 data will help to derive the best grain composition thanks to SIGMA and 3D radiative transfer modeling.

   \begin{figure}[h]
  	\begin{center}
  			\includegraphics[scale=0.3]{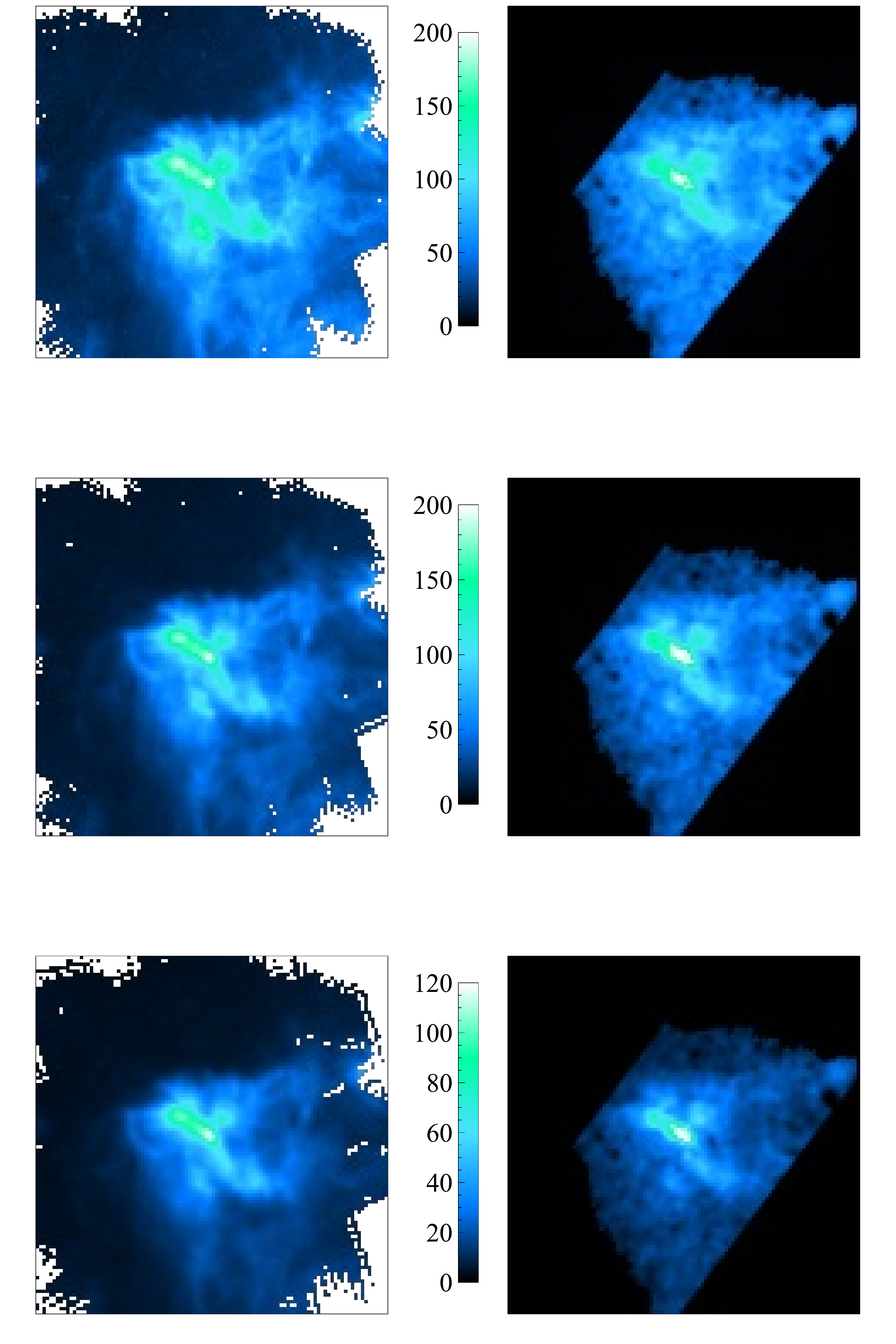}
  		\label{fig-Herschel}       
  		\caption{Herschel observations on left column at 250\,\mic, 350\,\mic, and 500\,\micb and corresponding modeled maps on right side using the dust model and density profile built from MIR scattering \cite{2016Lefevre}. Color scale is in MJy/sr, {box size is 40\arcmin}. The 500\,\micb modeled map shows a clear excess of emission towards the PSC suggesting that the dust population suitable to explain scattering observations is too emissive. }
  	\end{center}
  \end{figure}
  
\vspace{-1cm}
\subsection{SIGMA: a new tool to compute dust properties}
\label{sec-SIGMA}
 Dust grains evolve from bare simple elongated shape in the diffuse medium to large icy aggregates with irregular shapes inside dense cores. While dust properties in the diffuse medium are well-constrained, dust models able to explain multi-wavelength observations of dense environment including large grains are still lacking. Our goal was to develop a code able to reproduce dust properties of large aggregates in a fast computation time and to be able to relax free parameters like dust composition to better reproduce multi-wavelength observations.
 

SIGMA (Simple Icy Grain Model for Aggregates) relies on effective medium theory to compute refractive indexes from laboratory measurements and Mie theory applied to a distribution of hollow spheres (DHS) to mimic non–spherical dust shapes. An important characteristic of the DHS method is to reproduce the increased absorption opacity in the Rayleigh regime with respect to pure spherical shape. This emissivity enhancement is at least a factor of 2--3 compared to spherical shape (see Fig.~3, {left}). Also, the DHS method has been shown to be very effective in reproducing the properties of natural samples of complex particle shapes, including the correct positions of features related to solid-state resonances \cite[][and Fig.~3]{2001Fabian,2003Min,2011vanBreemen}. Attached to SIGMA, we also computed a set of dust size distributions representative of dynamical coagulation (see Fig.~3, right). Our goal is to characterize the dust evolution of the size distribution with temperature and density based on coagulation computation as a function of turbulence \cite{2011Ormel, 2013Hirashita}. SIGMA is able to deal with such size distributions including variable porosity as a function of size.

\begin{figure}[h]
	\label{fig-dust_properties}
	\begin{center}
		\begin{minipage}[b]{0.5\linewidth}
			\includegraphics[scale=0.27]{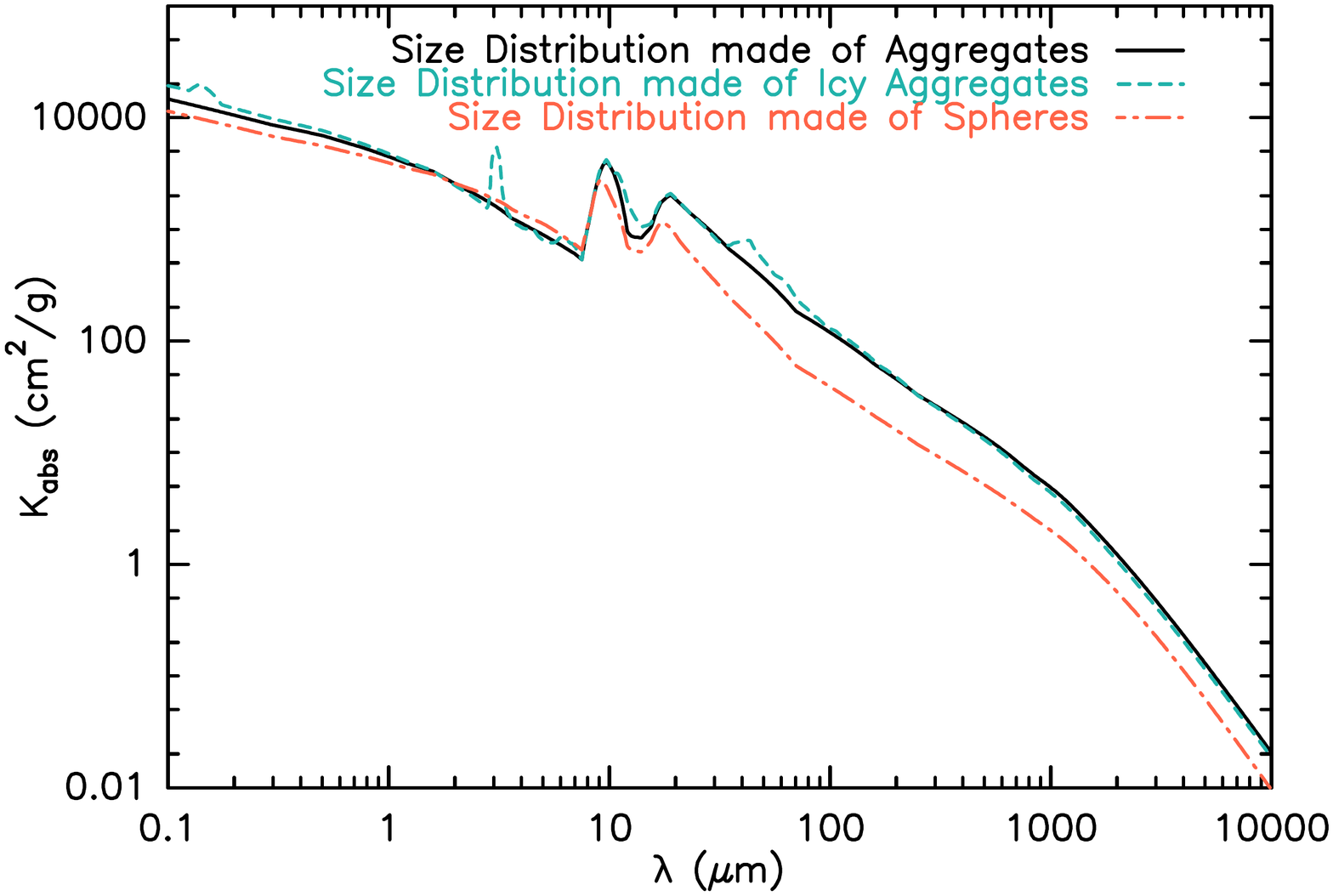}
		\end{minipage}
		\begin{minipage}[b]{0.45\linewidth}
			\includegraphics[scale=0.27]{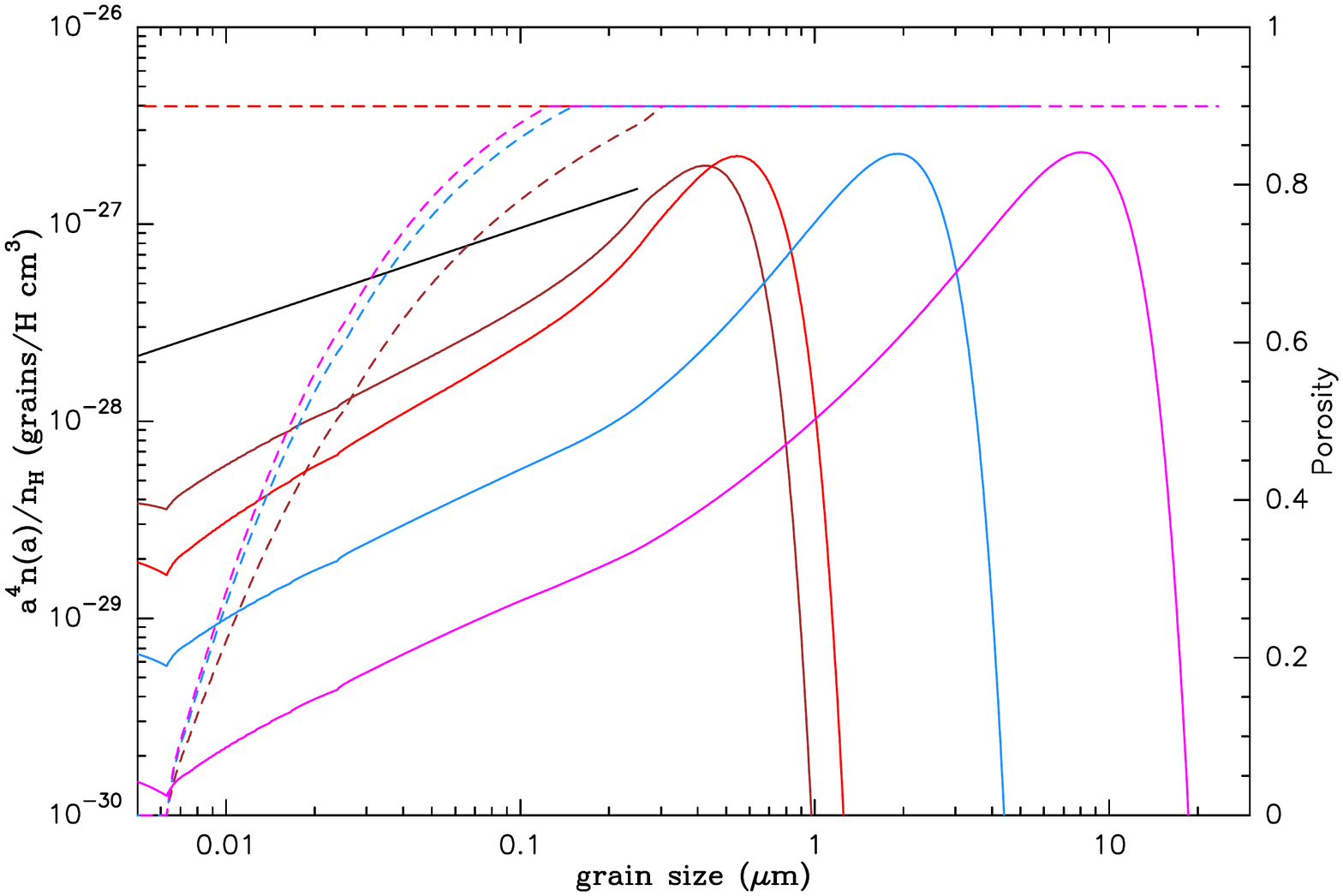}
		\end{minipage}
		\caption{\textbf{Left:} Dust emissivities for three dust models obtained with SIGMA: the first one made of a mixture of dust grains suitable to explain coreshine observations of L183 \cite{2016Min,2016Lefevre} in black, the same model with spherical shape instead of aggregates to show the enhancement of the emissivity in orange, and the last one including thin ice mantles (24\% in volume) in blue. \textbf{Right:} Dust size distribution evolution (solid lines) representative of a cloud contraction to reach 10$^6$\,H.cm$^{-3}$. We start from standard diffuse medium size distribution (in black). Red solid line describes the dust size distribution after 0.5\,My of evolution at constant porosity of 90\% \cite{2013Hirashita}, while brown curve takes into account that we start from non porous grains. Finally, blue and magenta curves are the evolution at non constant porosity after 2\,My and 5\,My, respectively. Porosity as a function of size for the different dust distributions are represented with dashed lines, 90\% corresponding to the compaction limit \cite{2011Ormel}.}
			
	\end{center}
\end{figure}

\vspace{-0.3cm}
\section{NIKA2 observations}
\label{sect-NIKA2}

   \begin{figure}[h!]
	\begin{center}
		\includegraphics[scale=0.4]{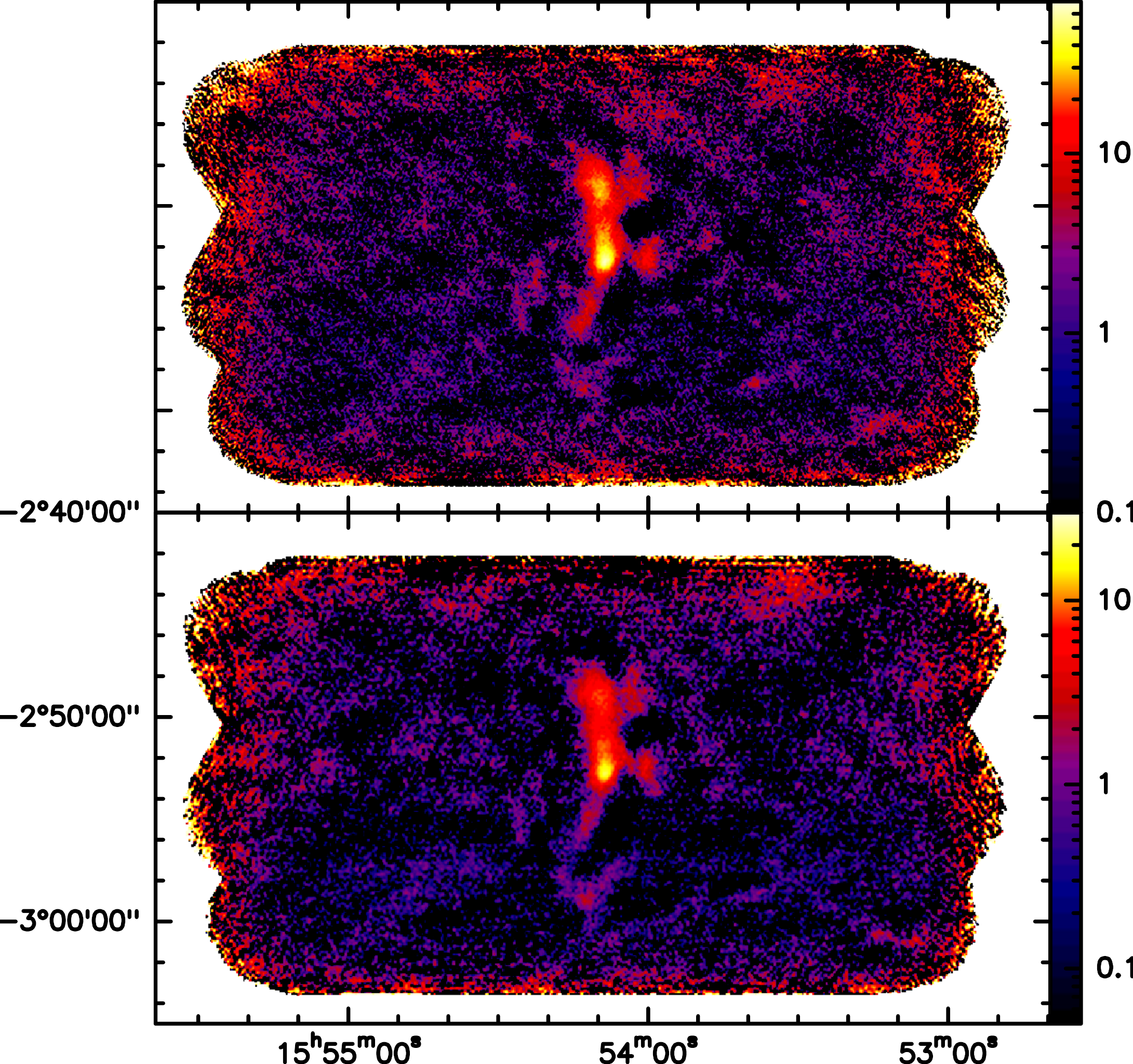}
		\label{fig-NIKA2}       
		
		\caption{NIKA2 maps of L183 at 1.15\,mm (upper panel) and 2\,mm (bottom panel) {obtained with $\sim$\,8 hours on source}. Data reduction was done with PIIC, color scale is in mJy/beam. }
	\end{center}
\end{figure}

L183 is about the Moon size while L134 is slightly smaller (see Fig.~5). We mapped L183 on $\sim$\,{450}\,arcmin$^2$ divided in two strips to keep the total integration time relatively short for each strip. We divided each strip in 3 on the fly (OTF) maps of $\sim$\,1000\,s each with a mapping speed of 60\arcsec\,s$^{-1}$. One OTF map is along the main direction of the strip and the two others with scanning direction $\pm$30\deg\ from the main direction to avoid artifacts (see Fig.~5). The average conditions during the observations were $\tau_{225}$\,$\sim$\,0.14 corresponding to a precipitable water vapor (PWV)\,$\sim$3\,mm. While the degradation of the integration time going from 2 to 4\,mm of PWV {is only 50 \%}, the
sky noise can become a real problem with high PWV for very extended maps of weak emission. L183 observations were completed in 2019 (see Fig.~4), but L134 observations are still on-going. We observed  last missing data of L134 in N$_2$H$^+$ with EMIR on early July 2018, and started to build a consistent cloud model of L134. Three over four of the PSCs are detected with our current NIKA2 observations but one is interestingly still not yet detected (L134N) despite its coreshine brightness comparable to others (see Fig.~5).

   \begin{figure}[h!]
	\begin{center}
		\begin{minipage}[b]{0.5\linewidth}
		\includegraphics[scale=0.3]{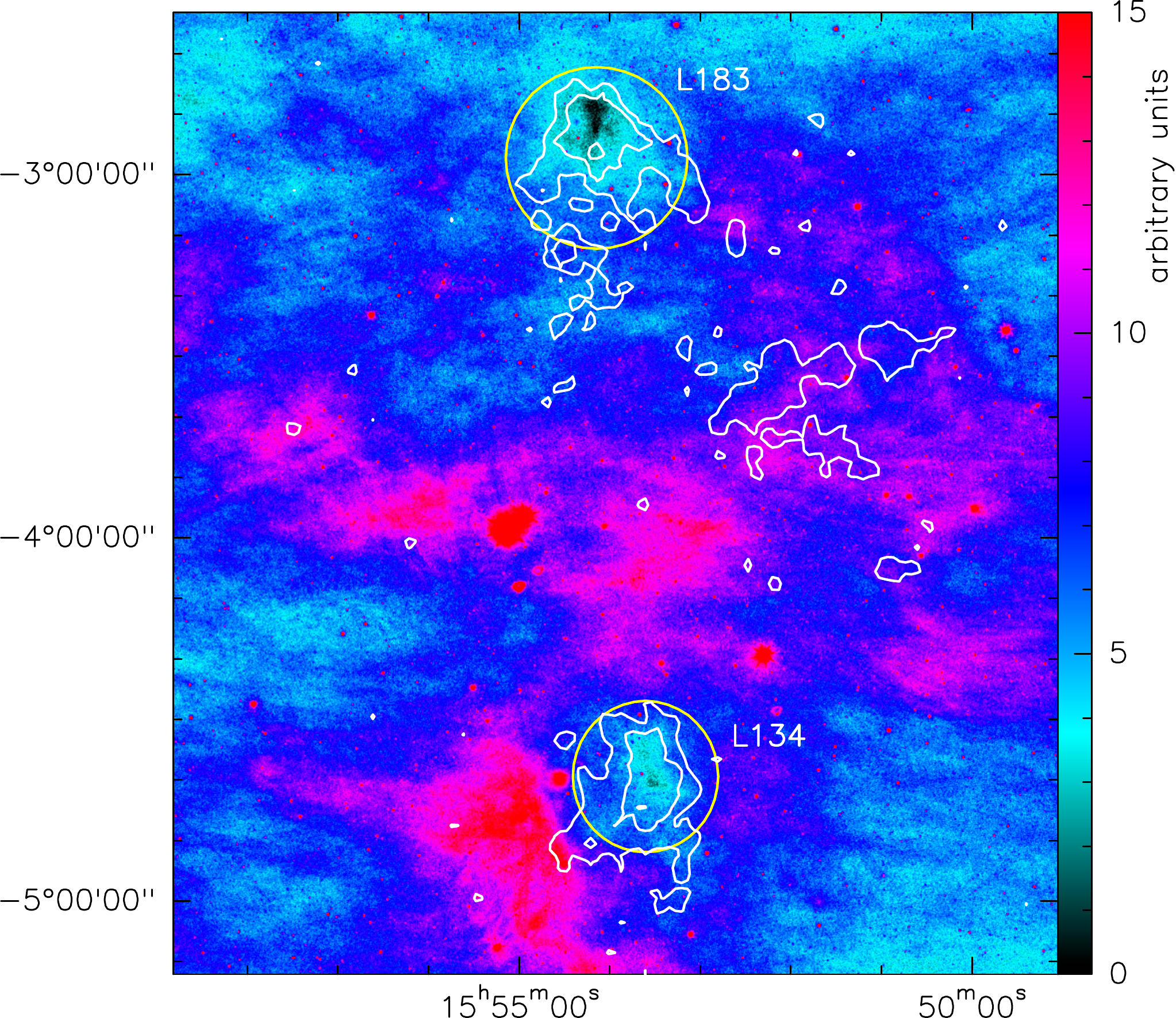}
		\end{minipage}
		\begin{minipage}[b]{0.4\linewidth}
		\includegraphics[scale=0.25]{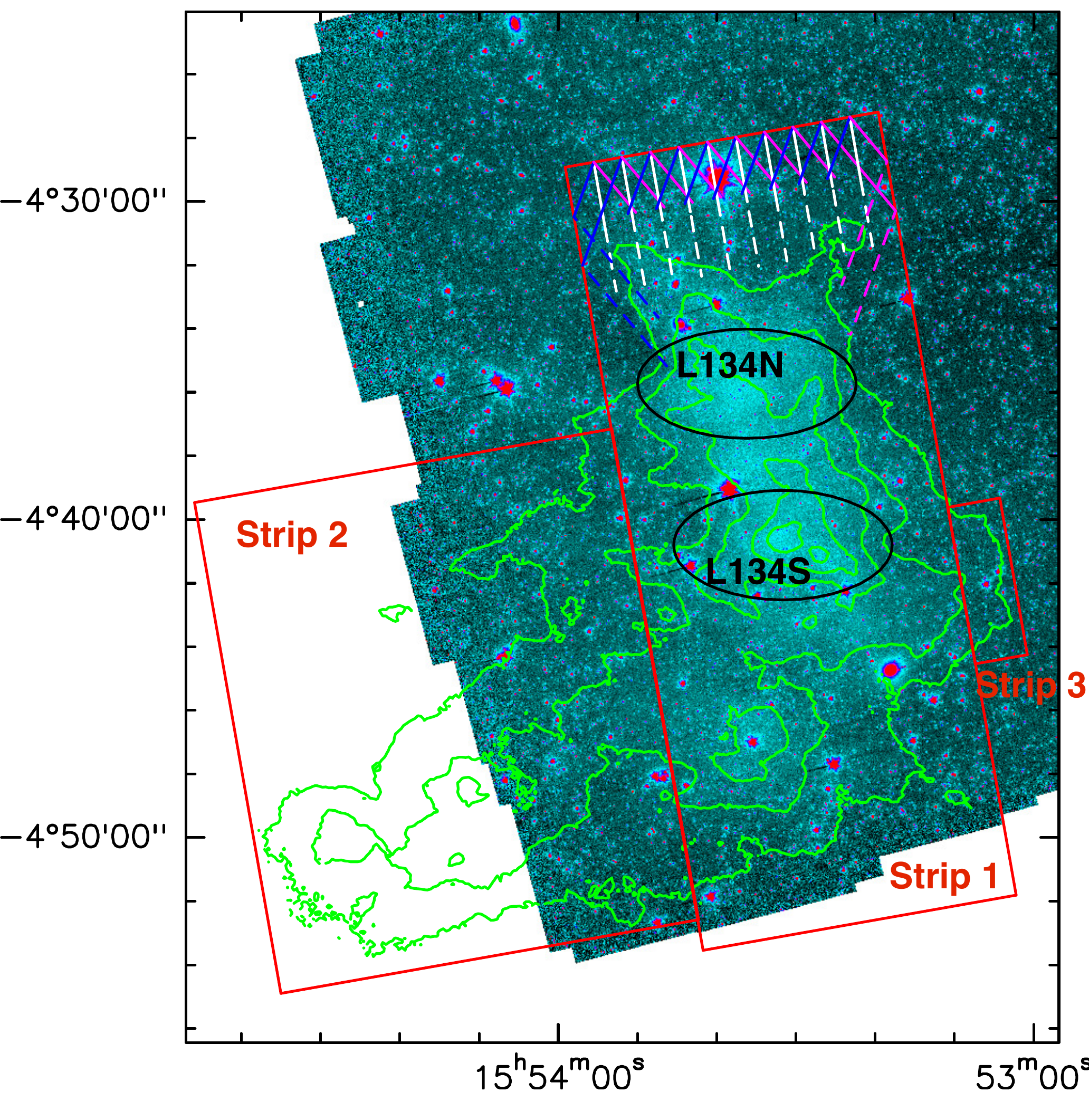}
		\end{minipage}

		\label{fig-L134-Spitzer}       
		
		\caption{\textbf{Left:} WISE extinction image of L183 and L134 at 12\mic. The yellow circles represent the external limits inside which the NIKA2 maps will be performed (30\arcmin for L183, $\sim$25\arcmin for L134). \textbf{Right:} Spitzer IRAC 1 (3.6\,\mic) image of L134 with Herschel SPIRE 250\,\mic\ contours superimposed. The diffuse emission in cyan is the coreshine effect. The black ellipses indicate the position of the depleted core (L134N, no C$^{18}$O, strong NH$_3$ and N$_2$H$^+$ emission) and of the undepleted core (L134S, C$^{18}$O emission). The three red boxes trace the regions to be mapped with NIKA2 ($\sim$ 465 arcmin$^2$ in total). When the box is elongated we will use three sweeping methods: along the main axis (white) or with 30\deg\,angles (blue and magenta) to limit artifacts.}
	\end{center}
\end{figure}

\noindent First, we will revisit the dust content of L183 thanks to NIKA2 data. In particular, we aim at:
\begin{itemize}
	\item{better constraining dust composition (silicates and ice mantles)}
	\item{find a suitable solution to match MIR and mm observations}
	\item{revisit dust size distributions made of aggregates mixing not only 2 components but making use of our coagulated dust size distributions.}
\end{itemize}
 By also modeling L134, when the observations will be done, we will better constrain the coagulation threshold for grain growth and study grain properties dependency with temperature and density. Our ultimate goal would be to infer typical emissivities that could be used beyond this study for example at larger scales (e.g. NIKA2 Large Program GASTON, see Peretto et al. proceeding in this book). Nevertheless, we do not exclude that grain properties vary from one region to the other making the extrapolation difficult. Performing NIKA2 observations of several PSCs with different environmental conditions will be the next step to distinguish between common dust evolution properties and individual peculiarities. In any case, NIKA2 observations will allow to validate dust models and characterize dust evolution in dense regions.

%

\begin{thebibliography}{21}
%
%

\bibitem{2013Pagani}
{Pagani}, L., {Lesaffre}, P., {Jorfi}, M. et al., A\&A 551, A38 (2013)
\bibitem{2017Kortgen}
{K{\"o}rtgen}, B., {Bovino}, S., {Schleicher}, Dominik R.~G., et al., MNRAS, 469, 2602 (2017)
\bibitem{2006Mouschovias}
Mouschovias, T., Tassis, K., Kunz, M., ApJ 646, 1043 (2006)
\bibitem{2010Pagani}
Pagani, L., Steinacker, J., Bacmann, A., et al., Science, 329, 1622 (2010)
\bibitem{2014Lefevre}
Lefèvre, C., Pagani, L., Juvela, M., et al., A\&A, 572, A20 (2014)
\bibitem{2016Lefevre}
Lefèvre, C., Pagani, L., Min, M., et al, A\&A, 585, L4 (2016)
\bibitem{2017SPHERE}
	Garufi, A.; Benisty, M.; Stolker, T., et al., The Messenger, 169, p. 32-37 (2017)
\bibitem{2015Pagani}
Pagani, L., Lefèvre, C., Juvela, M., et al., A\&A, 574, L5 (2015)
\bibitem{2019Villenave}
{Villenave}, M. and {Benisty}, M. and {Dent}, W.~R.~F., et al., A\&A, 624, A7 (2019)
\bibitem{2015aJuvela}
Juvela, M., Ristorcelli, I., Marshall, D. J., et al., A\&A, 584, A93, (2015a)
\bibitem{2015bJuvela}
Juvela, M., Demyk, K., Doi, Y., et al., A\&A, 584, A94, (2015b)
\bibitem{2019Zucker}
{Zucker}, C., {Speagle}, J., {Schlafly}, E. F. , ApJ, 879, 125 (2019)
\bibitem{2007Pagani}
Pagani, L., Bacmann, A., Cabrit, S., \& Vastel, C., A\&A, 467, 179, (2007)
\bibitem{2016Min}
Min, M., Rab, C., Woitke, P., Dominik, C., \& Menard, F., A\&A, 585, A13 (2016)
\bibitem{2017aDemyk}
Demyk, K., Meny, C., Lu, X. H., et al., A\&A, 600, A123, (2017a)
\bibitem{2017bDemyk}
Demyk, K., Meny, C., Leroux, H., et al., A\&A, 606, A50, (2017b)
\bibitem{2001Fabian}
Fabian, D., Henning, T., Jäger, C., et al., A\&A, 378, 228 (2001)
\bibitem{2003Min}
Min, M., Hovenier, J. W., \& de Koter, A., A\&A, 404, 35 (2003)
\bibitem{2011vanBreemen}
van Breemen, J. M., Min, M., Chiar, J. E., et al., A\&A, 526 (2011)
\bibitem{2011Ormel}
Ormel, C. W., Min, M., Tielens, A. G. G. M., et al., A\&A, 532 (2011)
\bibitem{2013Hirashita}
Hirashita, H. \& Li, Z. Y., MNRAS, 434, L70 (2013)
\end{thebibliography}
%
%

\end{document}